\documentclass[11pt]{article}
\usepackage[english]{babel}
\usepackage{graphicx}
\usepackage{amsmath}
\usepackage{amssymb}

\def\aj{AJ}					
\def\apj{ApJ}					
\def\apjl{ApJL}					
\def\apjs{ApJS}					
\def\mnras{MNRAS}				
\def\prd{Phys.~Rev.~D}				
\def\nat{Nature}				
\def\jcap{JCAP}			
\def\nar{NAR}			
\def\pasa{PASA}			

\newcommand{\HI}{{{\rm H}{\sc I}\,\,}}

\newcommand{\hmpc}{\,h^{-1}\,{\rm Mpc}}

\newcommand{\hgpcc}{\,h^{-3}\,{\rm Gpc}^3}

\newcommand{\mpc}{\,{\rm Mpc}}
\newcommand{\mpch}{\,h\,{\rm Mpc^{-1}}}

\newcommand{\degsq}{\, \, {\rm deg^{2}}}

\begin{document}

\begin{center}
{\bf \LARGE Probing the nature of dark energy through galaxy redshift surveys with radio telescopes}

\bigskip\bigskip
\large
Alan R. Duffy$^{}$\footnote{E-mail:~\textsf{mail@alanrduffy.com}}

\bigskip\bigskip
$^1${\small\it School of Physics, University of Melbourne, Parkville, VIC 3010, Australia}
\end{center}

\begin{abstract}
Galaxy redshift surveys using optical telescopes have, in combination with other cosmological probes, enabled 
precision measurements of the nature of dark energy. We show that radio telescopes are rapidly becoming competitive with optical facilities
in spectroscopic surveys of large numbers of galaxies. Two breakthroughs are driving this change. Firstly, individual radio telescopes 
are more efficient at mapping the sky thanks to the large field-of-view of new phased-array feeds. Secondly, ever more dishes can be correlated
in a cost-effective manner with rapid increases in computing power.
The next decade will see the coming of age of the $21 \rm \, cm$ radio wavelength as a cosmological probe as first the Pathfinders then, ultimately, the 
Square Kilometre Array is constructed. The latter will determine precise 3D positions for a billion galaxies, 
mapping the distribution of matter in the Universe over the last 12 billion years. This radio telescope will be able to constrain the equation of state of dark energy, 
and its potential evolution, to a precision rivalling that of future optical facilities such as DESI and Euclid.
\end{abstract}

\section{Introduction}\label{sec:intro}
The nature of dark energy, as parameterised through the ratio of the pressure to density in the equation of state $w$, can be probed through 
measurement of the distribution of matter on large scales, quantified by the matter power spectrum~\cite{SeoEisenstein}. 
Measuring the power spectrum across redshift, $z$, can constrain evolution in the equation of state, $w(z)$. At higher redshifts ($z{>>}1$) the impact
of dark energy becomes increasingly difficult to measure as the matter component dominates the evolution of the Universe. As distinct
from this `late' dark energy is an `early' dark energy (e.g.~\cite{Doran:06}) which has a much larger equation of state at early times such that its density was still
significant (and hence potentially measurable, or at least strongly constrained~\cite{Reichardt:12}) at high redshift.
In general, any result showing $w{\ne}-1$ or evolution in redshift would falsify the current concordance model of dark energy as a Cosmological Constant. 
Observationally we measure the positions of galaxies, constraining the galaxy power spectrum $P_{\rm gal}$
which traces the underlying matter power spectrum $P$. The 3D power spectrum $P_{\rm gal}$ is constructed  
from the angular distribution of these galaxies, together with their line of sight distance as given by their redshift. 
Historically, this redshift has been measured from the optical spectrum of the galaxy.

As well as measuring the shape of the matter power spectrum across a variety of scales it is possible to use a particular feature in the power spectrum known as the 
Baryonic Acoustic Oscillation (BAO) to further constrain cosmological parameters~\cite{Blake:11b}. This BAO appears as an enhancement of galaxies separated by ${\sim}105 \hmpc$.
The BAO is a remanent feature from sound waves oscillating in the Early Universe, which are left `frozen' in place as the sound speed 
of the coupled-photon-plasma stalls (at the baryon drag epoch $z{\approx}1020$) shortly after the Universe becomes fully neutral, known as the Epoch of Recombination ($z{\approx}1090$). 
This results in a region of increased density that will then collapse to form galaxies, separated by a preferred scale~\cite{Bashinsky:01, Bashinsky:02, Eisenstein:07a}. 
The exact scale of the BAO depends on the sound horizon at the Epoch of Recombination~\cite{EisensteinHu:98}. This horizon (to first order) depends on the physical matter 
and baryon densities; $\Omega_{m} h^2$ and $\Omega_{b} h^2$ respectively (where $h{\equiv}H_{0} / \rm 100 km/s/Mpc$ and $H_{0}$ is the locally measured Hubble parameter).
Ultimately the scale of the BAO acts as a `standard ruler' which we can measure at low redshift, and compare to the same scale at high redshifts (the highest redshift is seen
measured in the CMB; Cosmic Microwave Background). Modifications to the scale at late times constrains the accelerating expansion of the Universe, which we can attribute to dark energy.

In addition to measuring the matter power spectrum and BAO signal there are numerous other tests that can be performed using a galaxy redshift survey. 
For example, the validity of the theory of General Relativity on large scales can be tested through the rate of growth of structure~\cite{Blake:11a}.
As well as using BAOs as `standard rulers', one can also use `standard spheres'~\cite{Blake:11c} in the Alcock-Pazynski test~\cite{Alcock:79} 
to measure cosmic expansion history. The importance of measuring the power spectrum alongside these other tests 
was emphasised in~\cite{Signe:12}, with errors on constraints such as the total neutrino mass halving when the matter power spectrum 
was included with a dataset containing BAOs, CMB experiments and local Hubble parameter measurements.
In addition, by surveying significant spatial volumes the matter-radiation equality epoch can be probed directly in the power spectrum by
measuring the `turn-over' in the spectrum~\cite{Poole:13}.

To date, all of the cosmological parameter constraints from the matter power spectrum have been calculated from optical galaxy redshift surveys.
An alternative approach to determining redshifts with optical spectra is to use a naturally occurring spectral feature of cold gas, that of the spin-flip transition in neutral hydrogen ($\HI$) 
at a rest-frame wavelength of ${\sim}21 \rm \, cm$. Most $\HI$ resides in galaxies making it a ubiquitous tracer of large scale structures. It suffers little absorption or obscuration and naturally provides 
the spectral redshift making it an ideal tool to undertake a galaxy redshift survey~\cite{Abdalla:05}. The major drawback in using this line is 
that it is incredibly weak and requires enormous collecting area relative to optical facilities to enable a survey of a significant number of galaxies to be completed in a reasonable timeframe.

It has been suggested (e.g.~\cite{McQuinn:06, Mao:08a, Wyithe:08a, Wyithe:08b}) that radio telescopes are uniquely powerful probes of the high-redshift ($1.5{\le}z{\le}12$) matter power spectrum 
through constraining the large-scale $21 \rm \, cm$ fluctuations (with the signal averaged over scales of order 10$\mpc$).
This is a particularly powerful probe of the acoustic peak which would be typically an order of magnitude larger than smoothing scale of the observation, permitting a simple and direct
measure of the acoustic scale since the Epoch of Reionisation~\cite{Wyithe:08b}.
Furthermore, it has been argued~\cite{Mao:08b} that one can marginalise over the rapidly changing fluctuations of the $21 \rm \, cm$ signal during the Epoch of Reionisation
and recover cosmological parameter constraints almost as tight as if the underlying matter power spectrum had been directly measured.
Interestingly, measurements of the $21 \rm \, cm$ signal at these high-redshifts are particularly effective in constraining non-Gaussianity in the primordial density field to test inflation models~\cite{Joudaki:11}
as well as testing the nature of dark matter~\cite{Kadota:13}. 
We will restrict ourselves in this work to a like-for-like comparison between optical and radio galaxy (spectroscopic) redshift surveys
in the overlap region $z{<}2$ but note that the potential power of new cosmological probes with radio telescopes could outperform the more standard methodology discussed here.

A significant difference between next generation radio telescopes and optical facilities is that the latter typically build fewer, larger (and therefore more expensive) mirrors
while radio telescopes can be constructed from numerous, smaller mass-produced (therefore cheaper) dishes. A major expense in radio observations is in combining (correlating) 
the signal from multiple dishes, which is computationally intensive. Fortunately the cost of computing approximately halves every 18 months (a trend known as `Moore's Law, e.g.~\cite{Voller:02})
and offers a path to constructing extremely large telescopes that can detect the weak $\HI$ signal in a modest survey time.
An exciting further development in radio telescopes is their increasing field-of-view ($1-30\degsq$) with the development of phased array 
feeds~\cite{Chippendale:10}. These phased array feeds sit at the focus of the radio telescope and are analogous to pixels in a digital camera; enabling multiple beams across the sky
to be imaged at once.
This enables an enormous increase in survey speed (area of sky scanned to a given luminosity limit in a set amount of time) permitting the possibility of all sky surveys to heretofore unrivalled depth.
Taken together, new `front-end' feeds and ever cheaper computational `back-end' correlators would appear to suggest that the radio facilities will become competitive with optical telescopes 
in the coming decades. 

We introduce the quantitative measure of a galaxy redshift survey in constraining cosmological parameters, the effective volume, in Section~\ref{sec:veffest}. We then 
discuss the sample of galaxy redshift surveys we have compiled in Section~\ref{sec:surveys}. From this sample we discuss the rapid growth in the effective volume probed by galaxy redshift
surveys in Section~\ref{sec:scaling} and conclude in Section~\ref{sec:conclusion}.

\begin{table}
\centering
	\caption{  
	\label{tab:survey_list}  
	Comparison of effective volume of completed and planned spectroscopic surveys, optical (radio) wavelengths in the top (bottom) panel. 
  	For planned surveys we present the estimated completion date of the telescope and, in parenthesis, the duration of the survey.}

  \begin{tabular}[]{lcccc}
  Survey & $N_{\rm gal}$ & $V_{\rm eff}$ & (Est) Year & Ref \\
   & & $\hgpcc$ & (+ Survey) & \\
  \hline
  2dF GRS & 221414 & 0.08 & 2003 & \cite{Cole:05} \\
  SDSS Main DR3 &  374767 & 0.15 & 2004 & \cite{Tegmark:04a,Eisenstein:05a} \\
  SDSS LRG DR3 &  46748 & 0.4 & 2004 & \cite{Eisenstein:05a} \\
  6dFGS & 125071 & 0.08 & 2006 & \cite{Jones:09,Beutler:11} \\
  WiggleZ & 238000 & 1 & 2011 & \cite{Blake:10} \\
  SDSS CMASS &  283330 & 0.99 & 2012 & \cite{Anderson:12} \\
  SDSS BOSS & 1.5e6 & 2.8 & 2014 & \cite{Eisenstein:11a} \\
  TAIPAN &  4e5 & 0.23 & 2015(+3) & \cite{Beutler:11} \\
  DESI 14k & 24e6 & 6 & 2018(+4) & \cite{Mostek:12} \\
  Euclid & 1e8 & 19.7 & 2019(+6) & \cite{Amendola:13} \\
  \hline
   & & &  & \\
  \hline
  HIPASS & 4315 & 5e-4 & 2002 & \cite{Zwaan:05a} \\
  ALFALFA & 3e4 & 3e-3 & 2012 & \cite{Martin:10} \\
  ASKAP &  6e5 & 0.07 & 2016(+3) & \cite{Duffy:12b} \\
  FAST-19 & 2.6e6 & 0.23 & 2016(+2) & \cite{Duffy:08a} \\
  SKA1 & 4.4e7 & 2.95 & 2021(+1) & This work \\  
  SKA2 & 1e9 & 14 & 2025(+1) & \cite{Abdalla:05} 
  \end{tabular}
\end{table}

\section{Cosmological Parameter Estimation}\label{sec:veffest}
The ability of a galaxy redshift survey to constrain cosmological parameters depends on how well it has measured the matter power spectrum $P(k)$ at a given scale of interest or wavenumber $k$.
The typical measurement of the power spectrum will suffer from two sources of error. 
The first is sample variance, that not all $k$ modes are measured by a survey (or equivalently that it isn't sampling a representative volume of the Universe), 
the second is shot-noise in the measurements that are made on an individual mode. As given in~\cite{FKP,Tegmark1997}
the total error $\sigma_{\rm P}$ on the measurement of the power spectrum, $P(k,z)$, for a given $k$ with logarithmic bin width $\Delta(\log k)$ can be expressed as
\begin{eqnarray}\label{eqn:Perror} 
\left( \frac{\sigma_{P}}{P} \right)^{2}=2\frac{1}{4 \pi k^{3} \Delta(\log k)} \frac{(2\pi)^{3}}{V_{\rm eff}(k)} \left( \frac{1+nP}{nP} \right)^{2}\,,
\end{eqnarray}
where $P=P(k,z)$ and $n=n(z)$ is the number density of galaxies which are detected (making $nP$ dimensionless)
\begin{eqnarray}
n(z)=\int_{M_{\rm lim}(z)}^{\infty}{dN\over dVdM}\,dM\,,
\end{eqnarray}
in which $M_{\rm lim}(z)$ is the limiting $\HI$ mass threshold that can be detected for a given $\HI$ survey,
and $V_{\rm eff}(k)$ is the effective survey volume probed for a particular $k$-mode 
\begin{eqnarray}
\label{eqn:veff}
V_{\rm eff}(k)=\Delta \Omega \int_{0}^{\infty} \left(\frac{nP}{1+nP}\right)^{2} \frac{dV}{dzd\Omega}(z) dz\,,
\end{eqnarray}
in a survey of sky area surveyed $\Delta \Omega$ with $dV/dzd\Omega$ the infinitesimal comoving volume element for a Friedmann-Robertson-Walker universe. 
In Table~\ref{tab:survey_list}  we have calculated this effective volume for a number of surveys presented in Section~\ref{sec:surveys}. 

In designing a cosmological survey there is thus a balance between the precision at which a given mode is measured (through increased 
integration time spent on one region) and the number of modes measured by a survey through scanning larger areas of sky. The former approach 
of limiting a survey volume can increase the number of $k$ modes measured in the survey by increasing the galaxy number density to measure 
smaller scales (higher $k$-modes). However, non-linear effects of structure growth which are difficult to model at ever smaller scales 
typically limit the useful cosmological information to $k{\le}0.1 \mpch$~\cite{Beutler:11} in the local universe.

As noted by~\cite{Kaiser:86} for a survey of fixed number count the minimum error on the power spectrum measurement is attained by increasing
the volume surveyed such that a number density of sources $n\sim P^{-1}$ is achieved for the $k$-mode of interest. 
Therefore, the most cosmological information is achieved by maximising the survey volume (at a given number density) in which case a survey 
should prioritise sky area over depth as the former increases linearly in time while the latter increases only as the square root of the integration time~\cite{Duffy:08a}.
As discussed by~\cite{SeoEisenstein} designing a survey such that $nP$ is greater than unity is preferable in practice. For $nP>1$ 
the signal-to-noise per pixel in a map is increased enabling higher-order statistics to be calculated (such as testing for non-Gaussianity). Additionally, subsamples of the galaxy population
can be created to test for systematics. To that end, when we calculate the effective volume for radio surveys in Section~\ref{sec:effvol} we will use the recommended $nP{=}3$~\cite{SeoEisenstein}. 
We note that using the effective volume as a comparison between surveys is only a rough (30\%;~\cite{Eisenstein:05a}) measure for the statistical performance in
constraining cosmological parameters from the power spectrum, with error reducing as $\sigma_{\rm P} \propto V_{\rm eff}^{-1/2}$~\cite{Abdalla:05}.

\subsection{Bias}
The power spectrum as measured by a galaxy survey is not the same as the underlying matter power spectrum. Instead, the galaxies represent a 
biased tracer of the `true' distribution. We can relate the two power spectra through $P_{\rm gal}(k,z)=b^2 P(k,z)$ where 
$b$ is the bias parameter and can be both scale and redshift dependent. Particular surveys will tend to sample a population of 
galaxies with a given bias, for example, the SDSS Large Red Galaxy survey in~\cite{Eisenstein:05a} has $b{\approx}3$ which can be compared 
with the smaller blue star forming galaxies of~\cite{Beutler:11} in the 6dFGS with $b{\approx}2$. This has a significant effect on the 
detectability of the power spectrum as a high bias will raise the signal-to-noise of the detection for a given galaxy number count. 

The near unbiased (indeed potentially anti-biased) value of $\HI$ selected objects ($b{\le}1$~\cite{Basilakos:07}) may appear to be a mark against its use as a selection technique.
However, a low bias indicates a galaxy population that predominantly avoids regions of large overdensity, i.e. galaxy clusters, as the $\HI$
is removed through, for example, or tidally harassment by neighbours~\cite{Moore:98} or ram pressure stripping~\cite{Gunn:72} / strangulation~\cite{Larson:80} by the hot halo of the high-density regions.
The end result is that a blind $\HI$ survey will naturally avoid regions of large overdensity, and therefore the peculiar velocity effects inherent 
to these regions (such as the fingers of god). These effects are a significant limitation in probing the underlying matter power spectrum
to ever smaller scales (larger $k$-modes) as complex non-linear evolution of $P_{\rm gal}$ has to be modelled. By selecting
against these regions the $\HI$ survey (or any technique that selects a low bias tracer) 
will be able to probe higher $k$-modes still in the linear regime~\cite{Beutler:12} at low redshifts.
This will be a significant gain~\cite{Beutler:12,Duffy:12b} over current optical surveys which have a maximum of ${\sim}0.1 \mpch$~\cite{Beutler:11} in the local universe. We note that
for surveys at higher redshifts the linear regime extends to larger $k$-modes, for example WiggleZ~\cite{Blake:11b} and SDSS BOSS~\cite{Beutler:13} 
are able to probe to ${\sim}0.2\mpch$.

Extending the measured range of the matter power spectrum will maximise the cosmological constraint from a given survey and is particularly valuable 
for measuring neutrino masses~\cite{Signe:12}.
Probing to high $k$-modes is crucial for the technique of redshift space distortions which is not included in the discussion here, 
but which is valuable both for constraining dark energy and testing the validity of the theory of General Relativity on cosmological scales (e.g.~\cite{Beutler:12}).

\section{Survey List}\label{sec:surveys}
The last decades have seen a revolution in our capability to survey ever larger regions of space in a given time.
We have compiled a list of a number of these notable redshift surveys both at optical and radio wavelengths.
Although the list is not meant to be exhaustive, it is meant to be an instructive comparison between radio and optical facilities. To that end
we consider only spectroscopic redshift surveys. Notably, this means we exclude the upcoming photometric redshift 
Dark Energy Survey (DES~\cite{DES}), which is expected to find $10^{8}$ galaxies. 
In addition, we don't include the Large-scale Synoptic Survey Telescope (LSST;~\cite{Ivezic:08}) which will potentially discover billions of photometric redshifts.

At radio wavelengths we have ignored the possibilities of intensity mapping. As discussed in Section~\ref{sec:intro} this 
is a promising technique that would survey the sky at low angular resolution, 
co-adding the confused signal from numerous individual galaxies to measure the $\HI$ distribution on large scales~\cite{Battye:13}. We note a new single dish
facility (BINGO;~\cite{Battye:12}) has been proposed to undertake just such a survey, 
which could constrain the BAO scale to 2.4\% (and thereby the equation of state of dark energy to 16\%).

Another promising technique with radio telescopes is to detect galaxies by their continuum emission. This is significantly easier to detect than the $\HI$
signal ensuring many more galaxy detections for a given survey time but unfortunately these will not have spectroscopic redshifts. Instead, either the angular
power spectrum (distribution across the sky) can be measured, or cross-correlations with other catalogues can be used to statistically measure the line of sight
power spectrum. Continuum surveys are potentially a powerful cosmological probe which suffer different systematics to existing methods~\cite{Raccanelli:12}. Using
the upcoming Australian Square Kilometer Array (SKA) Pathfinder continuum survey EMU~\cite{Norris:11} 
together with Northern hemisphere counterparts LOFAR (the LOw Frequency ARray for radio astronomy~\cite{Rottgering:03})   
and WODAN (the Westerbork Observations of the Deep Apertif Northern sky survey~\cite{Rottgering:11}) the dark energy equation of state could be measured~\cite{Raccanelli:12} 
to ${\sim}5\%$ and evolution $\sigma(w_a){\sim}12\%$.

In comparing galaxy redshift surveys we have recalculated the quoted effective volumes to the same fiducial scale, $k=0.065\mpch$ which is approximately the position of the first BAO
`peak' in the power spectrum~\cite{Duffy:08a}. For completeness where possible we have quoted the known (or estimated) dark energy equation of state constraints from the survey,
but due to the inherent difficulty of various dataset combinations used by different authors and cosmological parameter choices (although we have attempted to be as consistent
as possible) we choose to compare the surveys using their effective volume. As noted by~\cite{Abdalla:05} the 
error on the equation of state should approximately decrease as the square root of this volume.

In Table~\ref{tab:survey_list} we list the various surveys, their effective volumes, number of galaxies found (or estimated to be detected) 
and the (estimated) year of completion. For upcoming facilities we note the expected completion data plus the estimated time taken for the
given survey (in parenthesis). 

\subsection{Optical Surveys}
The first optical survey we consider is the 2-degree Field (2dF) Galaxy Redshift Survey~\cite{Colless:01,Colless:03} which successfully measured the 
matter power spectrum using $10^5$ galaxies~\cite{Cole:05}.
The two-point correlation function (the Fourier transform of the power spectrum) was measured using a similar number of galaxies~\cite{Tegmark:04a} 
from the Sloan Digital Sky Survey (SDSS~\cite{York:00}). We note that the measured matter power spectrum of~\cite{Tegmark:04a} from the SDSS Main Galaxy sample 
in Data Release 3 is in fact only half the effective volume of that survey, making it equivalent to the 2dF GRS, and instead we take the full volume as 
described in~\cite{Eisenstein:05a}. In addition, the Baryonic Acoustic Oscillation was detected in a subsample of Large-Red Galaxies~\cite{Eisenstein:05a} from this dataset.
The resultant error on the equation of state for the SDSS DR3 Main and LRG sample (which uses the BAO measurement and CMB constraints~\cite{WMAP3}) 
is approximately 20 per cent~\cite{Eisenstein:05a}. 

The SDSS has continued to survey the Universe, with several data releases (e.g.~\cite{Abazajian:09,Schlegel:09}) 
adding substantial numbers of new galaxies over ever larger cosmological volumes (e.g.~\cite{Kazin:10,Percival:10}). 
From these updates we have selected the 
SDSS III - Baryon Oscillation Spectroscopic Survey (BOSS;~\cite{Dawson:13}) and its first release the SDSS CMASS sample of~\cite{Anderson:12}.
We have recalculated the quoted $V_{\rm eff}(k=0.1\mpch) = 2.2 {\rm Gpc}^3$ of the SDSS CMASS result to $V_{\rm eff}(k=0.065\mpch) = 0.99 \hgpcc$.
The SDSS CMASS sample, together with CMB results from the Wilkinson Microwave Anisotropy Probe Year 7 (WMAP7;~\cite{Komatsu:11}) data release, was able 
to constrain a non-evolving equation of state (at $z\sim0.57$) to approximately $27\%$~\cite{Anderson:12}, and with the addition of SDSS LRG constraints from~\cite{Padmanabhan:12a}
this tightens to $20\%$. The full BOSS sample is expected~\cite{Dawson:13} to constrain an evolving dark energy equation of state
\footnote{$w(a)=w_{p} + (a_{p} - a)w'$ where $w_{p}$ is the equation of state at a pivot scale factor $a_{p}=2/3$} to $1\sigma$ errors of $\sigma(w_{p})= 0.030$ and $\sigma(w')= 0.32$.

The WiggleZ Dark Energy Survey~\cite{Drinkwater:10} obtained spectroscopic redshifts of $2.3\times10^5$ galaxies at large cosmological distances ($0.2{<}z{<}1$)
which were used to measure the matter power spectrum~\cite{Blake:10}. We note that, as described in Section~\ref{sec:intro}, this survey in particular has been 
successful in utilising a variety of cosmological probes such as redshift space distortions, and growth of structure tests, to further constrain cosmological parameters.
However, for this work we only consider the ability of the WiggleZ Dark Energy Survey to measure the matter power spectrum~\cite{Parkinson:12}. 
Using the power spectrum measurement, together with WMAP7 CMB data and a local measure of the Hubble expansion~\cite{Riess:09} to anchor this high redshift 
galaxy survey, WiggleZ DES measured a non-evolving equation of state to $8$ per cent~\cite{Parkinson:12}. 

Currently, the leading optical constraints of the matter power spectrum in the {\it local} Universe is from the 6 degree Field Galaxy Survey (6dFGS;~\cite{Jones:04,Jones:09}). 
This project utilised the large field-of-view of the UK Schmidt Telescope to survey a full hemisphere (i.e. half the sky) providing the most precise 
measurement of the local matter power spectrum~\cite{Beutler:12}. In~\cite{Beutler:11} it was shown that adding 6dFGS to the SDSS LRG sample with the WMAP7 CMB 
dataset was able to improve the constraints on the dark energy equation of state by 24 per cent (from 17 per cent uncertainty to 13).
We were unable to find the expected dark energy constraints for the TAIPAN\footnote{TAIPAN: Transforming Astronomical Imaging surveys through Polychromatic Analysis of Nebulae} survey, but based on the estimated performance from~\cite{Beutler:11} who found it to be
similar to the WALLABY\footnote{http://www.atnf.csiro.au/research/WALLABY} 
survey on the Australian SKA Pathfinder~\cite{Johnston:08, Deboer:09}, we can quote the radio telescope performance of constraining a non-evolving
equation of state to 20 per cent~\cite{Duffy:12b, Duffy:12c} for TAIPAN.

The future for optical spectroscopic surveys will be, from the ground, DESI (Dark Energy Spectroscopic Instrument~\cite{Abdalla:12,Levi:13}) and, in space, the Euclid satellite~\cite{Amendola:13}.
DESI is forecast to measure~\cite{Levi:13} as many as 18 million emission-line galaxies, 4 million LRGs and 3 million quasars across a fiducial $14000\degsq$ of sky and a 
redshift range ($0{<}z{<}3.5$). We take the effective volume calculated by~\cite{Mostek:12}, formally for the BigBOSS mission but as discussed in~\cite{FontRibera:13}
these values are representative for DESI.
Euclid aims to detect redshifts of hundreds of millions of galaxies~\cite{Amendola:13} during the six year mission, 
along with additional constraints through the weak gravitational lensing signature of intervening mass concentrations amongst other tests.
For these new surveys we have taken the Fisher-matrix error prediction of~\cite{FontRibera:13} who performed a large cosmological parameter search, using
Planck priors, and various combinations of optical surveys. In particular, we quote their Table XIII for the case of Planck plus DESI which would constrain 
the dark energy equation of state $w_{0}$ and its evolution $w'$ (according to $w=w_{0}+(1-a)w'$) to $1\sigma$ errors of $0.10$ and $0.33$ (using the full power spectrum to $k{\le}0.1\hmpc$,
and BAO). For the case of Euclid,~\cite{FontRibera:13} conservatively assumed only half the number of galaxies as predicted by~\cite{Amendola:13} would be detected. When~\cite{FontRibera:13} 
added the Euclid results to the existing Planck plus DESI errors they found only a marginal improvement in the errors to $\sigma(w_0)=0.099$ and $\sigma(w')=0.29$.

\subsection{Radio Surveys}
The first large volume $\HI$ survey we consider is the HI Parkes All Sky Survey (HIPASS;~\cite{Meyer:04,Zwaan:04}) which performed a full
southern hemisphere survey with the Parkes telescope (effective surface area ${\sim}10^4 \rm \,m^2$) and detected $4\times10^3$ 
galaxies. A smaller sky area but deeper redshift survey, the Arecibo Legacy Fast ALFA survey~\cite{Giovanelli:05}, used the much larger 
Arecibo dish (${\sim}10^5 \rm \,m^2$) to discover $3\times10^4$ galaxies~\cite{Martin:10}. 
Looking ahead the largest single dish in the world, the Five-hundred metre Aperture Spherical Telescope~\cite{Nan:06, Nan:11},
will potentially find $10^6$ galaxies in its current 19-beam system configuration~\cite{Duffy:08a}. 
In~\cite{Duffy:08a} a future upgrade to a 100-beam system was proposed which could detect ${\sim}10^7$ galaxies,
and constrain a non-evolving equation of state to five per cent (when combined with Planck priors). The fiducial
19-beam system has four times lower effective volume, resulting in constraints a factor $\sqrt{4}$ worse~\cite{Abdalla:05} than this proposed
100-beam system. This would therefore result in FAST-19 being competitive with the constraints of~\cite{Beutler:11} who used 6dFGS (together with the SDSS LRG sample and CMB data from WMAP7).

In analogy with the optical Euclid facility, the future for radio-led cosmological surveys will be the Square Kilometre Array, with a fiducial telescope footprint area of $10^6 \rm \,m^2$ to be constructed by 2025.
There are several smaller precursor facilities for the SKA that are nearing completion in the meantime. 
The first is the low-frequency Murchison Widefield Array~\cite{Lonsdale:09} which will survey the high redshift Universe, in particular the Epoch of Reionisation, 
but is unlikely to be used to constrain the matter power spectrum as detailed here. 
A limited sky area but extremely deep galaxy redshift survey in $\HI$ will instead be carried out by the 60-dish Meer-Karoo Array Telescope (MeerKAT;~\cite{Booth:09}) and 
a larger sky area but shallower survey by the 36-dish ASKAP~\cite{Johnston:08, Deboer:09}.
In particular, the latter facility will potentially detect~\cite{Duffy:12b,Duffy:12c} $7\times10^5$ galaxies out to $z=0.26$ that will enable
measurements of the matter power spectrum to constrain a non-evolving dark energy equation of state (when combined with Planck priors) to better than 20 per cent,
as mentioned before this is comparable to TAIPAN.

After the SKA pathfinder is the Phase 1 SKA, termed SKA1, which has approximately 10 per cent of the final collecting area of the `full' SKA (termed SKA phase 2, or SKA2). 
We have used the straw-man design specifications\footnote{www.skatelescope.org/uploaded/21705\_130\_Memo\_Dewdney.pdf}  for this telescope.
In particular, we assume that 70\% of the 250-15m dishes lie within 2.5km of the core, with a system temperature of 30K (as given for the 1-2GHz frequency coverage of the 
survey redshift range $z=0-0.42$ we consider here). We note that extending the redshift range to $z=1$ increases the effective volume by less than ten per cent 
as the number density of galaxies beyond $z=0.42$ rapidly falls off (for the assumed integration time). We consider a fiducial integration time of
8 hours (for the angular resolution of the telescope this will maximise galaxy counts and volume surveyed~\cite{Duffy:08a}). 
Sky coverage is assumed to be the same as ASKAP at $3\Pi$ steradian. Errors on the equation of state approximately decrease as $V_{\rm eff}^{1/2}$~\cite{Abdalla:05}
such that SKA1 could determine a non-evolving equation of state to several per cent.

Our quoted numbers for the SKA2 are from the model of~\cite{Abdalla:05} of an all-sky survey with a field-of-view, that increases as the square of the (redshift) wavelength and is $1\degsq$ at $z=0$,
detecting approximately $10^9$ galaxies over the course of a year. Such a survey could 
potentially constrain the dark energy equation of state $w$ to one percent and evolution over several redshift bins up to $z<1.5$ parameterised as $w=w_{0} + w_{1}z$
with $w_0$ to 3.5 per cent, and the error on the evolution $w_1$ to 10 per cent~\cite{Abdalla:05}.

\section{Effective Volume Scaling Relations}\label{sec:scaling}
\begin{figure}
  \includegraphics[width=\columnwidth]{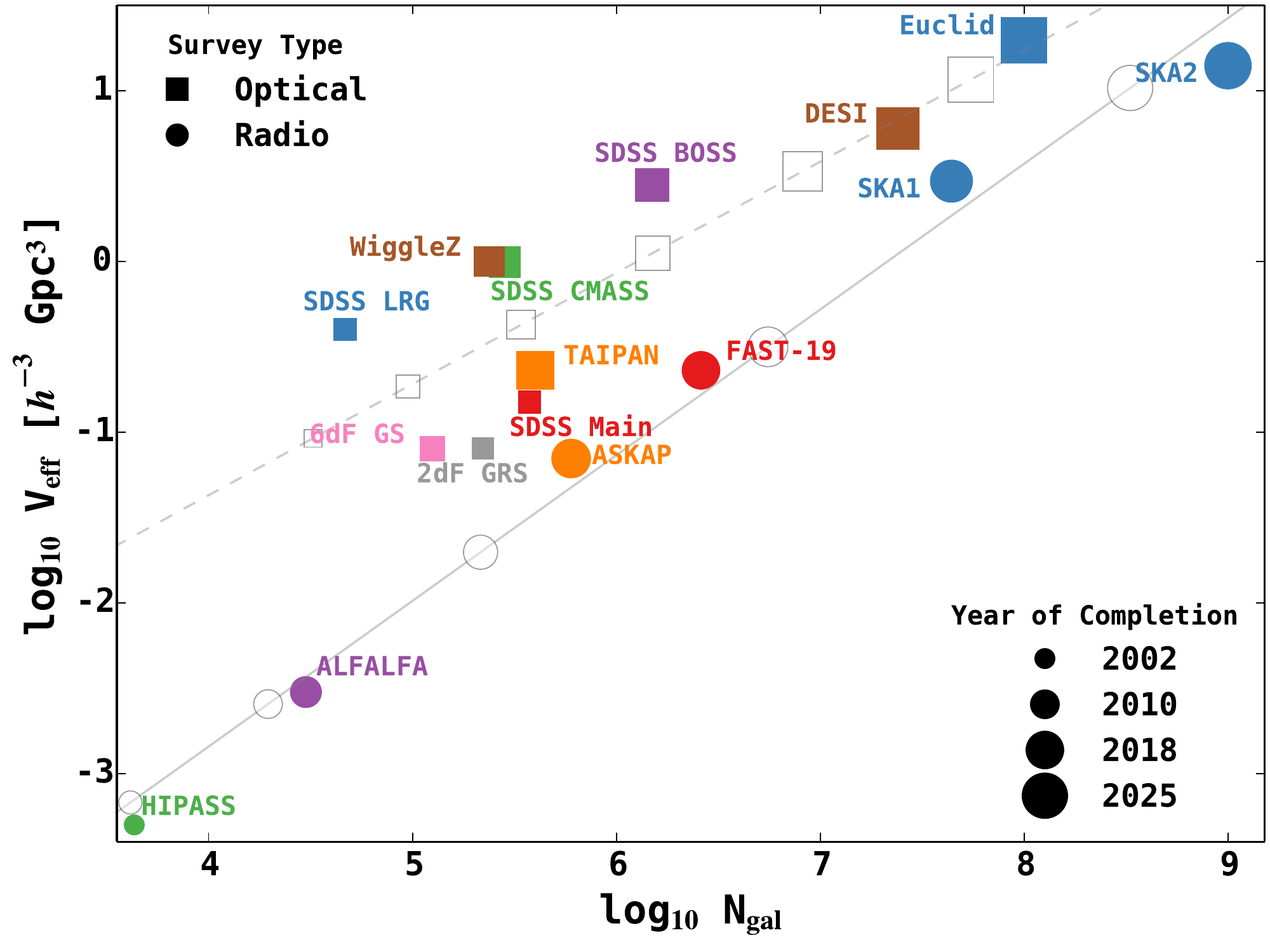}%
  \caption{\label{fig:veffngal} The effective volume of a survey (Eqn.~\ref{eqn:veff}) as a function of total galaxy count for various galaxy redshift
  surveys at optical and radio wavelengths. Symbol size indicates the approximate year that the survey finished (will finish). The 
  optical (radio) surveys are given by squares (circles) with colours indicating the survey (also annotated beside each point). 
  We fit a power law scaling between the effective volume of the survey and the number of galaxies (eqn.~\ref{eqn:veffngal}) 
  which we show as the grey solid (dashed) curve for radio (optical) surveys.
  We have overlaid the scaling of effective volume with survey completion year (Eqn.\ref{eqn:vefftime}) as the grey square (circle) symbols on the 
  dashed (solid) curve for the optical (radio) survey, respectively.}
\end{figure}

In Fig.~\ref{fig:veffngal} we show the effective volumes of the various surveys listed in Table~\ref{tab:survey_list} as a function of the 
number of galaxies they have found / are expected to find. It is clear that there has been a rapid increase in the size of the surveys over the last decade yet 
these successes will be dwarfed by the future surveys proposed by SKA and Euclid. 

\subsection{Effective Volume and Galaxy Counts}\label{sec:effvol}
From a low base relative to optical surveys the effective volume probed by the radio telescopes as well as the number of galaxies detected 
is predicted to rapidly increase.
Indeed, there appears to be a strong correlation between the volume and number of galaxies found across all surveys in Fig.~\ref{fig:veffngal}. 
This correlation is by design, as it naturally follows from the diminishing return with increasing the observed number density of galaxies significantly beyond the optimal
measurement $n{\sim}P^{-1}$ of Eqn.~\ref{eqn:veff}, as noted by~\cite{Kaiser:86,SeoEisenstein}. As discussed in Section~\ref{sec:veffest}, to gain the largest effective 
volume for a given number of galaxies, a survey should maximise the volume it probes at a constant minimum number density
meaning the effective volume will scale approximately linearly with the number of galaxies detected in that volume. In fact, the scaling will
be sub-linear as the majority of galaxies are detected at lower redshifts than the maximum redshift at which the minimum number density is achieved. This
results in $nP>>1$ for the modes within the survey, which is a suboptimal use of the galaxies in constraining the power spectrum (ideally the survey would have a constant
number density to the edge of the survey volume). In practice, when we calculated the radio surveys in this work we chose the conservative measurement goal of $nP\sim3$
as argued by~\cite{SeoEisenstein} and discussed in Section~\ref{sec:veffest}.

To determine if this simple sub-linear scaling occurs we fit each wavelength with a power law, shown as a grey solid (dashed) curve for the radio (optical) survey, of the form
\begin{equation}\label{eqn:veffngal}
\log_{10} V_{\rm eff}(k=0.65\mpch) = A_{\rm gal} \log_{10} N_{\rm gal} + B_{\rm gal},
\end{equation}
where $\rm A_{gal}, B_{gal}$ are given by $0.85, -6.26, \, (0.65, -3.98)$ for radio (optical) respectively. As expected the effective volume is a sub-linear 
scaling with galaxy numbers detected. This correlation is of course by design as individual galaxy redshift surveys will maximise the effective volume
probed based on the galaxy sample they are targeting. We can see in Fig.~\ref{fig:veffngal} the SKA1 prediction has a higher effective volume
than the best fit power law (solid line), indicating that the galaxies will likely be used more efficiently than previous radio surveys have achieved. 
The SKA2 prediction lies below the best fit line indicating it will attempt to survey more galaxies than needed to probe the target effective volume, however,
there are of course other scientific cases to be made with such a large dataset which may justify the increased galaxy count. 

Note that for a given number of detected galaxies the optical surveys always probe a larger effective volume than the radio survey. This is in part due to the larger bias that
optically derived galaxies exhibit relative to radio detections, which increases the given power spectrum normalisation at $k=0.065\mpch$ 
as the square of the bias. This means that the condition from~\cite{Kaiser:86} of $n{\sim}P^{-1}$ becomes $n{\sim}(Pb^2)^{-1}$, 
reducing the required number density for a given a measurement of the power spectrum by a factor $b^2$.

\subsection{Effective Volume in Time}
\begin{figure}
  \includegraphics[width=\columnwidth]{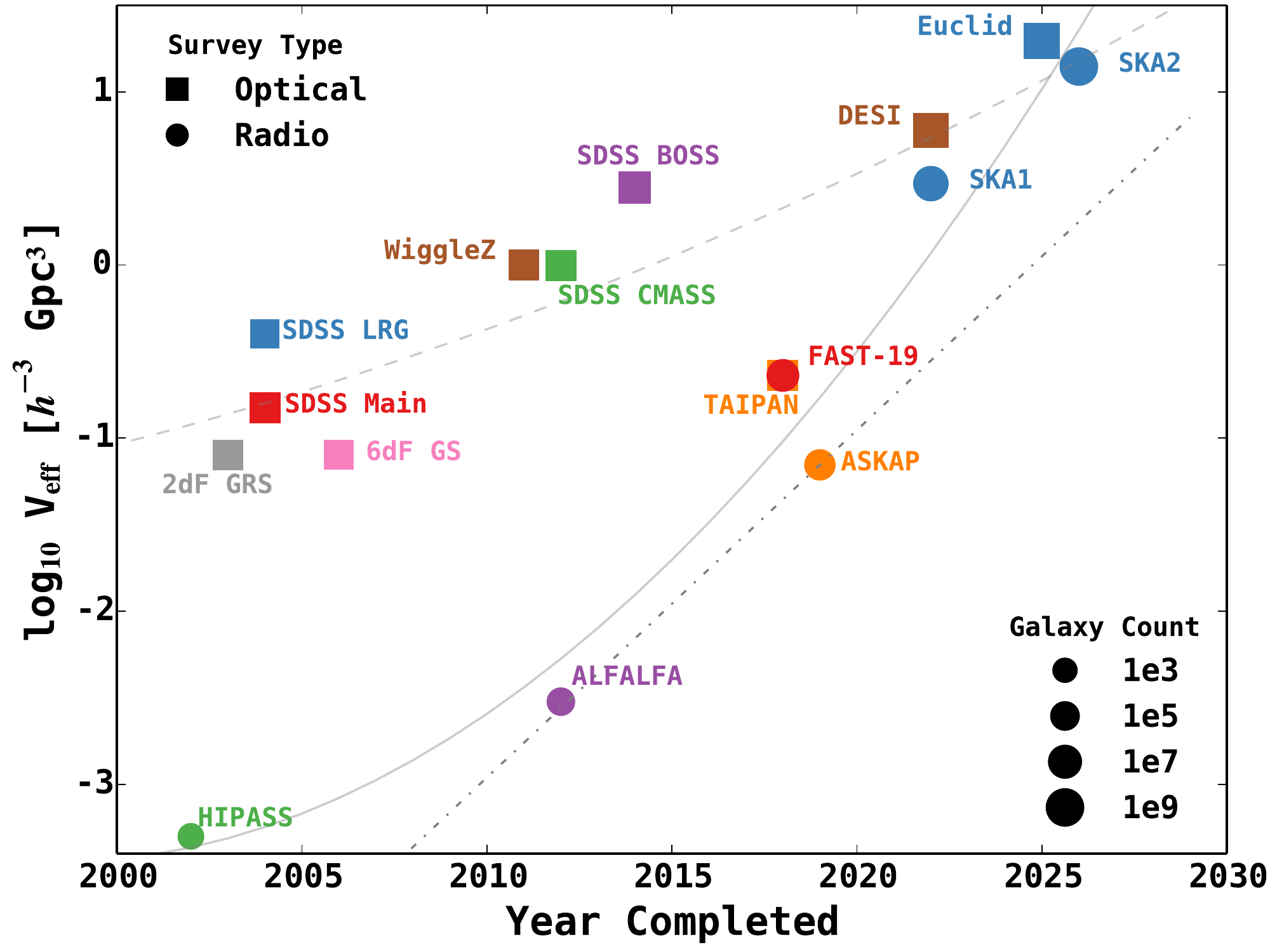}%
  \caption{\label{fig:vefftime} We plot the effective volume of a galaxy redshift survey (Eqn.~\ref{eqn:veff}) as a function of the year it was completed 
  for both optical and radio facilities. The rapid rise of the radio effective volumes has been due in part to ever cheaper
  computational hardware to efficiently, and cost-effectively, correlate ever larger numbers of dishes. The doubling of computing power every 18 months, 
  known as `Moore's Law' (e.g. Eqn. 1 of~\cite{Voller:02}),
  is shown as a grey dot-dash curve, in which we assume that a doubling of computer power corresponds to a doubling of effective volume. We see
  that from a low base the rapid rise of radio telescope performance is in fact predicted to outpace `Moore's Law'.}
\end{figure}

We consider the scaling of the effective volume probed by the surveys in time in Fig.~\ref{fig:vefftime}. We now find a log-linear 
relation between the typical effective volume at $k=0.65\mpch$ of the radio surveys and the year $T_{year}$ that the survey finished, given by
\begin{equation}\label{eqn:vefftime}
\begin{split}
\log_{10} V_{\rm eff}(k) & = \rm A_{time} (T_{year}-2000)^2 + \\
 & \rm B_{time} (T_{year}-2000) + C_{time},
\end{split}
\end{equation}
where the constants of proportionality $\rm A_{time}, B_{time}, C_{time}$ are $6.3\times10^{-3}, 2.1\times10^{-2}, -3.4 \, (1.1\times10^{-3}, 5.5\times10^{-2}, -1.0)$ for the radio (optical) surveys.
The exact best fitting values are arbitrary but the relative values indicate that the effective volumes probed by the radio surveys are rapidly advancing on the optical surveys.
This would suggest a `coming of age' of the radio wavelength as a cosmological probe, with parity to optical surveys occurring by 2025, with the completion of the full SKA.

The rapid rise in effective volume in time is indicative of another well known log-linear relation, in which computing power doubles every 18 months 
known as `Moore's Law' (e.g. Eqn. 1 of~\cite{Voller:02}). We have overlaid this relation as a grey dot-dash curve in Fig.~\ref{fig:vefftime} with arbitrary normalisation
(we have chosen the computationally challenging ASKAP facility as a reasonable example of a supercomputer powered `correlator'). We have implicitly assumed that doubling the 
computing power available will double the number of galaxies surveyed and, assuming a uniform distribution of galaxies, thus a doubling of effective volume probed.
It is clear that from a low base radio surveys will have to rapidly increase in time, indeed faster than `Moore's Law', to reach the expected SKA1 and SKA2 targets. 
Since `Moore's Law' can be thought of as a doubling of computing power every 18 months for {\it fixed} cost it then raises the question
of whether the computing power to analyse these surveys will be sufficient at current cost. If not, full correlation of all dishes for SKA1 / SKA2 
may be delayed until such time as computing power has caught up given current budget constraints.

\section{Conclusion}\label{sec:conclusion}
In this work we have compiled a representative sample of past, current and planned spectroscopic galaxy redshift surveys at radio and optical wavelengths. 
We have either taken the effective volume of the surveys from the literature or calculated this metric ourselves to enable a crude
comparison to be made between drastically different surveys covering a time span from 2002 - 2025.
We note that the error on the power spectrum (and hence cosmological parameter estimation) reduces as the square root of the 
effective volume (Eqn.~\ref{eqn:veff}) making this a convenient metric for comparison between surveys.

We discussed the various dark energy constraints from different surveys, noting that based on galaxy power spectrum measurements
the DESI and Euclid optical surveys are essentially equivalent~\cite{FontRibera:13} and of comparable constraining power 
to the expected full SKA~\cite{Abdalla:05}. Until this time however, the optical surveys will be vastly dominant over contemporary radio surveys. The exception
to this conclusion is the promising technique of intensity mapping~\cite{Wyithe:08a} with radio telescopes. This approach can rapidly survey the high redshift matter power
spectrum, while less useful to constrain `late' dark energy (as matter dominates over the dark energy component above $z{>}2$)
this would be invaluable for probing `early' dark energy which would have a significant, and measurable, dark energy signature at early times~\cite{Doran:06}.

As a general result we found that for a given galaxy count an optical survey would probe a larger effective volume than the corresponding radio survey, 
as shown in Fig.~\ref{fig:vefftime}. We attribute this to the higher typical bias $b$ of galaxies selected by an optical survey, which results in an increased
measured power spectrum at a given scale (increasing as $b^2$) and hence larger effective volume probed. 
A counterpoint to this is that a galaxy sample with low bias naturally selects against over-dense regions, which minimises the hard-to-model effects of
non-linear structure growth on the matter power spectrum at small scales (high $k$-modes).
A low bias survey can thus increase its ability to constrain cosmological parameters by extending the measured power spectrum to higher $k$-modes.

We considered the effective volume as a function of the year the survey was completed in Fig.~\ref{fig:vefftime} and found a strong log-linear relation.
The rapid rise of radio telescope performance means that despite a lower base in early 2000s the planned radio surveys will potentially reach parity with
optical surveys by 2025 (i.e. with the full SKA).
The log-linear growth of the survey performance was reminiscent of the doubling in computing power every 18 months, known as `Moore's Law'. However,
we note that the predicted rise of the radio telescopes is in fact significantly more rapid than `Moore's Law'. If computing power is the limiting factor in the SKA
design then this raises the possibility of a delay in full SKA2 by of order five years (or a significantly more expensive computing hardware outlay in 2025).
Alternatively, the community may be able to invest in radically more efficient software design to mitigate this issue.
Regardless, the next decade will see the development of wide field-of-view phased array feeds on numerous small dishes enabling rapid, cost-effective 
$\HI$ surveys of the sky. These surveys are naturally spectroscopic in nature and will result in 3D positions of potentially billions of galaxies, probing large effective volumes. 

The next generation of radio facilities will thus become `software' telescopes that can harness new computing power to remain competitive with other wavelengths, 
and represent a `coming of age' of radio telescopes in which they probe a similar effective volume to optical based surveys.


\begin{thebibliography}{[10]}

\bibitem{SeoEisenstein}
 \textsc{H.\,J. {Seo}} and  \textsc{D.\,J. {Eisenstein}} \jr{\apj}
  \textbf{598}(December), 720--740 (2003).


\bibitem{Doran:06}
 \textsc{M.~{Doran}} and  \textsc{G.~{Robbers}} \jr{\jcap} \textbf{6}(June), 26
  (2006).


\bibitem{Reichardt:12}
 \textsc{C.\,L. {Reichardt}},  \textsc{R.~{de Putter}},  \textsc{O.~{Zahn}},
  and  \textsc{Z.~{Hou}} \jr{\apjl} \textbf{749}(April), L9 (2012).


\bibitem{Blake:11b}
 \textsc{C.~{Blake}},  \textsc{T.~{Davis}},  \textsc{G.\,B. {Poole}},
  \textsc{D.~{Parkinson}},  \textsc{S.~{Brough}},  \textsc{M.~{Colless}},
  \textsc{C.~{Contreras}},  \textsc{W.~{Couch}} \etal{} \jr{\mnras}
  \textbf{415}(August), 2892--2909 (2011).


\bibitem{Bashinsky:01}
 \textsc{S.~{Bashinsky}} and  \textsc{E.~{Bertschinger}} \jr{Physical Review
  Letters} \textbf{87}(8), 081301 (2001).


\bibitem{Bashinsky:02}
 \textsc{S.~{Bashinsky}} and  \textsc{E.~{Bertschinger}} \jr{\prd}
  \textbf{65}(12), 123008 (2002).


\bibitem{Eisenstein:07a}
 \textsc{D.\,J. {Eisenstein}},  \textsc{H.\,J. {Seo}},  and
  \textsc{M.~{White}} \jr{\apj} \textbf{664}(August), 660--674 (2007).


\bibitem{EisensteinHu:98}
 \textsc{D.\,J. {Eisenstein}} and  \textsc{W.~{Hu}} \jr{\apj}
  \textbf{496}(March), 605--+ (1998).


\bibitem{Blake:11a}
 \textsc{C.~{Blake}},  \textsc{S.~{Brough}},  \textsc{M.~{Colless}},
  \textsc{C.~{Contreras}},  \textsc{W.~{Couch}},  \textsc{S.~{Croom}},
  \textsc{T.~{Davis}},  \textsc{M.\,J. {Drinkwater}},  \textsc{K.~{Forster}}
  \etal{} \jr{\mnras} \textbf{415}(August), 2876--2891 (2011).


\bibitem{Blake:11c}
 \textsc{C.~{Blake}},  \textsc{K.~{Glazebrook}},  \textsc{T.\,M. {Davis}},
  \textsc{S.~{Brough}},  \textsc{M.~{Colless}},  \textsc{C.~{Contreras}},
  \textsc{W.~{Couch}},  \textsc{S.~{Croom}} \etal{} \jr{\mnras}
  \textbf{418}(December), 1725--1735 (2011).


\bibitem{Alcock:79}
 \textsc{C.~{Alcock}} and  \textsc{B.~{Paczynski}} \jr{\nat}
  \textbf{281}(October), 358 (1979).


\bibitem{Signe:12}
 \textsc{S.~{Riemer--S{\o}rensen}},  \textsc{C.~{Blake}},
  \textsc{D.~{Parkinson}},  \textsc{T.\,M. {Davis}},  \textsc{S.~{Brough}},
  \textsc{M.~{Colless}},  \textsc{C.~{Contreras}},  \textsc{W.~{Couch}} \etal{}
  \jr{\prd} \textbf{85}(8), 081101 (2012).


\bibitem{Poole:13}
 \textsc{G.\,B. {Poole}},  \textsc{C.~{Blake}},  \textsc{D.~{Parkinson}},
  \textsc{S.~{Brough}},  \textsc{M.~{Colless}},  \textsc{C.~{Contreras}},
  \textsc{W.~{Couch}},  \textsc{D.\,J. {Croton}} \etal{} \jr{\mnras}
  \textbf{429}(March), 1902--1912 (2013).


\bibitem{Abdalla:05}
 \textsc{F.\,B. {Abdalla}} and  \textsc{S.~{Rawlings}} \jr{\mnras}
  \textbf{360}(June), 27--40 (2005).


\bibitem{McQuinn:06}
 \textsc{M.~{McQuinn}},  \textsc{O.~{Zahn}},  \textsc{M.~{Zaldarriaga}},
  \textsc{L.~{Hernquist}},  and  \textsc{S.\,R. {Furlanetto}} \jr{\apj}
  \textbf{653}(December), 815--834 (2006).


\bibitem{Mao:08a}
 \textsc{X.\,C. {Mao}} and  \textsc{X.\,P. {Wu}} \jr{\apjl}
  \textbf{673}(February), L107--L110 (2008).


\bibitem{Wyithe:08a}
 \textsc{J.\,S.\,B. {Wyithe}} and  \textsc{A.~{Loeb}} \jr{\mnras}
  \textbf{383}(January), 606--614 (2008).


\bibitem{Wyithe:08b}
 \textsc{J.\,S.\,B. {Wyithe}},  \textsc{A.~{Loeb}},  and  \textsc{P.\,M.
  {Geil}} \jr{\mnras} \textbf{383}(January), 1195--1209 (2008).


\bibitem{Mao:08b}
 \textsc{Y.~{Mao}},  \textsc{M.~{Tegmark}},  \textsc{M.~{McQuinn}},
  \textsc{M.~{Zaldarriaga}},  and  \textsc{O.~{Zahn}} \jr{\prd} \textbf{78}(2),
  023529 (2008).


\bibitem{Joudaki:11}
 \textsc{S.~{Joudaki}},  \textsc{O.~{Dor{\'e}}},  \textsc{L.~{Ferramacho}},
  \textsc{M.~{Kaplinghat}},  and  \textsc{M.\,G. {Santos}} \jr{Physical Review
  Letters} \textbf{107}(13), 131304 (2011).


\bibitem{Kadota:13}
 \textsc{K.~{Kadota}},  \textsc{Y.~{Mao}},  \textsc{K.~{Ichiki}},  and
  \textsc{J.~{Silk}} \jr{ArXiv e-prints}(December) (2013).


\bibitem{Voller:02}
 \textsc{V.\,R. {Voller}} and  \textsc{F.~{Port{\'e}-Agel}} \jr{Journal of
  Computational Physics} \textbf{179}(July), 698--703 (2002).


\othercit
\bibitem{Chippendale:10}
 \textsc{A.\,P. {Chippendale}},  \textsc{J.~{O'Sullivan}},
  \textsc{J.~{Reynolds}},  \textsc{R.~{Gough}},  \textsc{D.~{Hayman}},  and
  \textsc{S.~{Hay}},
{Phased array feed testing for astronomy with ASKAP},
 in: Phased Array Systems and Technology (ARRAY), 2010 IEEE International
  Symposium on, pp.648-652, 12-15 Oct. 2010,  (October 2010),  pp.\,648--652.


\bibitem{Cole:05}
 \textsc{S.~{Cole}},  \textsc{W.\,J. {Percival}},  \textsc{J.\,A. {Peacock}}
  \etal{} \jr{\mnras} \textbf{362}(September), 505--534 (2005).


\bibitem{Tegmark:04a}
 \textsc{M.~{Tegmark}},  \textsc{M.\,R. {Blanton}},  \textsc{M.\,A. {Strauss}},
   \textsc{F.~{Hoyle}},  \textsc{D.~{Schlegel}},  \textsc{R.~{Scoccimarro}},
  \textsc{M.\,S. {Vogeley}},  \textsc{D.\,H. {Weinberg}} \etal{} \jr{\apj}
  \textbf{606}(May), 702--740 (2004).


\bibitem{Eisenstein:05a}
 \textsc{D.\,J. {Eisenstein}},  \textsc{I.~{Zehavi}},  \textsc{D.\,W. {Hogg}},
  \textsc{R.~{Scoccimarro}},  \textsc{M.\,R. {Blanton}},  \textsc{R.\,C.
  {Nichol}},  \textsc{R.~{Scranton}},  \textsc{H.\,J. {Seo}} \etal{} \jr{\apj}
  \textbf{633}(November), 560--574 (2005).


\bibitem{Jones:09}
 \textsc{D.\,H. {Jones}},  \textsc{M.\,A. {Read}},  \textsc{W.~{Saunders}},
  \textsc{M.~{Colless}},  \textsc{T.~{Jarrett}},  \textsc{Q.\,A. {Parker}},
  \textsc{A.\,P. {Fairall}},  \textsc{T.~{Mauch}} \etal{} \jr{\mnras}
  \textbf{399}(October), 683--698 (2009).


\bibitem{Beutler:11}
 \textsc{F.~{Beutler}},  \textsc{C.~{Blake}},  \textsc{M.~{Colless}},
  \textsc{D.\,H. {Jones}},  \textsc{L.~{Staveley-Smith}} \etal{} \jr{\mnras}
  \textbf{416}(October), 3017--3032 (2011).


\bibitem{Blake:10}
 \textsc{C.~{Blake}},  \textsc{S.~{Brough}},  \textsc{M.~{Colless}},
  \textsc{W.~{Couch}},  \textsc{S.~{Croom}},  \textsc{T.~{Davis}},
  \textsc{M.\,J. {Drinkwater}},  \textsc{K.~{Forster}} \etal{} \jr{\mnras}
  \textbf{406}(August), 803--821 (2010).


\bibitem{Anderson:12}
 \textsc{L.~{Anderson}},  \textsc{E.~{Aubourg}},  \textsc{S.~{Bailey}},
  \textsc{D.~{Bizyaev}},  \textsc{M.~{Blanton}},  \textsc{A.\,S. {Bolton}},
  \textsc{J.~{Brinkmann}},  \textsc{J.\,R. {Brownstein}},  \textsc{A.~{Burden}}
  \etal{} \jr{\mnras} \textbf{427}(December), 3435--3467 (2012).


\bibitem{Eisenstein:11a}
 \textsc{D.\,J. {Eisenstein}},  \textsc{D.\,H. {Weinberg}},
  \textsc{E.~{Agol}},  \textsc{H.~{Aihara}},  \textsc{C.~{Allende Prieto}},
  \textsc{S.\,F. {Anderson}},  \textsc{J.\,A. {Arns}},
  \textsc{{\'E}.~{Aubourg}},  \textsc{S.~{Bailey}},  \textsc{E.~{Balbinot}},
  and  \textsc{et~al.} \jr{\aj} \textbf{142}(September), 72 (2011).


\othercit
\bibitem{Mostek:12}
 \textsc{N.~{Mostek}},  \textsc{K.~{Barbary}},  \textsc{C.\,J. {Bebek}},
  \textsc{A.\,T. {Dey}},  \textsc{J.~{Edelstein}},  \textsc{P.~{Jelinsky}},
  \textsc{A.\,G. {Kim}},  \textsc{M.\,L. {Lampton}} \etal{},
{Mapping the universe with BigBOSS},
 in: Society of Photo-Optical Instrumentation Engineers (SPIE) Conference
  Series, , Society of Photo-Optical Instrumentation Engineers (SPIE)
  Conference Series,  Vol.\,8446 (September 2012).


\bibitem{Amendola:13}
 \textsc{L.~{Amendola}},  \textsc{S.~{Appleby}},  \textsc{D.~{Bacon}},
  \textsc{T.~{Baker}},  \textsc{M.~{Baldi}},  \textsc{N.~{Bartolo}},
  \textsc{A.~{Blanchard}},  \textsc{C.~{Bonvin}},  \textsc{S.~{Borgani}}
  \etal{} \jr{Living Reviews in Relativity} \textbf{16}(September), 6 (2013).


\bibitem{Zwaan:05a}
 \textsc{M.\,A. {Zwaan}},  \textsc{M.\,J. {Meyer}},
  \textsc{L.~{Staveley-Smith}},  and  \textsc{R.\,L. {Webster}} \jr{\mnras}
  \textbf{359}(May), L30--L34 (2005).


\bibitem{Martin:10}
 \textsc{A.\,M. {Martin}},  \textsc{E.~{Papastergis}},
  \textsc{R.~{Giovanelli}},  \textsc{M.\,P. {Haynes}},  \textsc{C.\,M.
  {Springob}},  and  \textsc{S.~{Stierwalt}} \jr{\apj} \textbf{723}(November),
  1359--1374 (2010).


\bibitem{Duffy:12b}
 \textsc{A.\,R. {Duffy}},  \textsc{A.~{Moss}},  and
  \textsc{L.~{Staveley-Smith}} \jr{\pasa} \textbf{29}(May), 202--211 (2012).


\bibitem{Duffy:08a}
 \textsc{A.\,R. {Duffy}},  \textsc{R.\,A. {Battye}},  \textsc{R.\,D. {Davies}},
   \textsc{A.~{Moss}},  and  \textsc{P.\,N. {Wilkinson}} \jr{\mnras}
  \textbf{383}(January), 150--160 (2008).


\bibitem{FKP}
 \textsc{H.\,A. {Feldman}},  \textsc{N.~{Kaiser}},  and  \textsc{J.\,A.
  {Peacock}} \jr{\apj} \textbf{426}(May), 23--37 (1994).


\bibitem{Tegmark1997}
 \textsc{M.~{Tegmark}} \jr{Physical Review Letters} \textbf{79}(November),
  3806--3809 (1997).


\bibitem{Kaiser:86}
 \textsc{N.~{Kaiser}} \jr{\mnras} \textbf{222}(September), 323--345 (1986).


\bibitem{Basilakos:07}
 \textsc{S.~{Basilakos}},  \textsc{M.~{Plionis}},  \textsc{K.~{Kova{\v c}}},
  and  \textsc{N.~{Voglis}} \jr{\mnras} \textbf{378}(June), 301--308 (2007).


\bibitem{Moore:98}
 \textsc{B.~{Moore}},  \textsc{G.~{Lake}},  and  \textsc{N.~{Katz}} \jr{\apj}
  \textbf{495}(March), 139 (1998).


\bibitem{Gunn:72}
 \textsc{J.\,E. {Gunn}} and  \textsc{J.\,R. {Gott}III.} \jr{\apj}
  \textbf{176}(August), 1 (1972).


\bibitem{Larson:80}
 \textsc{R.\,B. {Larson}},  \textsc{B.\,M. {Tinsley}},  and  \textsc{C.\,N.
  {Caldwell}} \jr{\apj} \textbf{237}(May), 692--707 (1980).


\bibitem{Beutler:12}
 \textsc{F.~{Beutler}},  \textsc{C.~{Blake}},  \textsc{M.~{Colless}},
  \textsc{D.\,H. {Jones}},  \textsc{L.~{Staveley-Smith}},  \textsc{G.\,B.
  {Poole}},  \textsc{L.~{Campbell}},  \textsc{Q.~{Parker}},
  \textsc{W.~{Saunders}},  and  \textsc{F.~{Watson}} \jr{\mnras}
  \textbf{423}(July), 3430--3444 (2012).


\bibitem{Beutler:13}
 \textsc{F.~{Beutler}},  \textsc{S.~{Saito}},  \textsc{H.\,J. {Seo}},
  \textsc{J.~{Brinkmann}},  \textsc{K.\,S. {Dawson}},  \textsc{D.\,J.
  {Eisenstein}},  \textsc{A.~{Font-Ribera}},  \textsc{S.~{Ho}} \etal{}
  \jr{ArXiv e-prints}(December) (2013).


\bibitem{DES}
 \textsc{{The Dark Energy Survey Collaboration}} \jr{ArXiv Astrophysics
  e-prints}(October) (2005).


\bibitem{Ivezic:08}
 \textsc{Z.~{Ivezic}},  \textsc{J.\,A. {Tyson}},  \textsc{E.~{Acosta}},
  \textsc{R.~{Allsman}},  \textsc{S.\,F. {Anderson}},  \textsc{J.~{Andrew}},
  \textsc{R.~{Angel}},  \textsc{T.~{Axelrod}} \etal{} \jr{ArXiv e-prints}(May)
  (2008).


\bibitem{Battye:13}
 \textsc{R.\,A. {Battye}},  \textsc{I.\,W.\,A. {Browne}},
  \textsc{C.~{Dickinson}},  \textsc{G.~{Heron}},  \textsc{B.~{Maffei}},  and
  \textsc{A.~{Pourtsidou}} \jr{\mnras} \textbf{434}(September), 1239--1256
  (2013).


\bibitem{Battye:12}
 \textsc{R.\,A. {Battye}},  \textsc{M.\,L. {Brown}},  \textsc{I.\,W.\,A.
  {Browne}},  \textsc{R.\,J. {Davis}},  \textsc{P.~{Dewdney}},
  \textsc{C.~{Dickinson}},  \textsc{G.~{Heron}},  \textsc{B.~{Maffei}},
  \textsc{A.~{Pourtsidou}},  and  \textsc{P.\,N. {Wilkinson}} \jr{ArXiv
  e-prints}(September) (2012).


\bibitem{Raccanelli:12}
 \textsc{A.~{Raccanelli}},  \textsc{G.\,B. {Zhao}},  \textsc{D.\,J. {Bacon}},
  \textsc{M.\,J. {Jarvis}},  \textsc{W.\,J. {Percival}},  \textsc{R.\,P.
  {Norris}},  \textsc{H.~{R{\"o}ttgering}},  \textsc{F.\,B. {Abdalla}} \etal{}
  \jr{\mnras} \textbf{424}(August), 801--819 (2012).


\bibitem{Norris:11}
 \textsc{R.\,P. {Norris}},  \textsc{A.\,M. {Hopkins}},  \textsc{J.~{Afonso}},
  \textsc{S.~{Brown}},  \textsc{J.\,J. {Condon}},  \textsc{L.~{Dunne}},
  \textsc{I.~{Feain}},  \textsc{R.~{Hollow}} \etal{} \jr{\pasa}
  \textbf{28}(August), 215--248 (2011).


\bibitem{Rottgering:03}
 \textsc{H.~{R{\"o}ttgering}} \jr{\nar} \textbf{47}(September), 405--409
  (2003).


\bibitem{Rottgering:11}
 \textsc{H.~{R{\"o}ttgering}},  \textsc{J.~{Afonso}},  \textsc{P.~{Barthel}},
  \textsc{F.~{Batejat}},  \textsc{P.~{Best}},  \textsc{A.~{Bonafede}},
  \textsc{M.~{Br{\"u}ggen}},  \textsc{G.~{Brunetti}} \etal{} \jr{Journal of
  Astrophysics and Astronomy} \textbf{32}(December), 557--566 (2011).


\bibitem{Colless:01}
 \textsc{M.~{Colless}},  \textsc{G.~{Dalton}},  \textsc{S.~{Maddox}},
  \textsc{W.~{Sutherland}},  \textsc{P.~{Norberg}},  \textsc{S.~{Cole}},
  \textsc{J.~{Bland-Hawthorn}},  \textsc{T.~{Bridges}} \etal{} \jr{\mnras}
  \textbf{328}(December), 1039--1063 (2001).


\bibitem{Colless:03}
 \textsc{M.~{Colless}},  \textsc{B.\,A. {Peterson}},  \textsc{C.~{Jackson}},
  \textsc{J.\,A. {Peacock}},  \textsc{S.~{Cole}},  \textsc{P.~{Norberg}},
  \textsc{I.\,K. {Baldry}},  \textsc{C.\,M. {Baugh}} \etal{} \jr{ArXiv
  Astrophysics e-prints}(June) (2003).


\bibitem{York:00}
 \textsc{D.\,G. {York}},  \textsc{J.~{Adelman}},  \textsc{J.\,E.
  {Anderson}Jr..},  \textsc{S.\,F. {Anderson}},  \textsc{J.~{Annis}},
  \textsc{N.\,A. {Bahcall}},  \textsc{J.\,A. {Bakken}},
  \textsc{R.~{Barkhouser}},  \textsc{S.~{Bastian}} \etal{} \jr{\aj}
  \textbf{120}(September), 1579--1587 (2000).


\bibitem{WMAP3}
 \textsc{D.\,N. {Spergel}},  \textsc{R.~{Bean}},  \textsc{O.~{Dor{\'e}}}
  \etal{} \jr{\apjs} \textbf{170}(June), 377--408 (2007).


\bibitem{Abazajian:09}
 \textsc{K.\,N. {Abazajian}},  \textsc{J.\,K. {Adelman-McCarthy}},
  \textsc{M.\,A. {Ag{\"u}eros}},  \textsc{S.\,S. {Allam}},  \textsc{C.~{Allende
  Prieto}},  \textsc{D.~{An}},  \textsc{K.\,S.\,J. {Anderson}},  \textsc{S.\,F.
  {Anderson}},  \textsc{J.~{Annis}},  \textsc{N.\,A. {Bahcall}},  and
  \textsc{et~al.} \jr{\apjs} \textbf{182}(June), 543--558 (2009).


\bibitem{Schlegel:09}
 \textsc{D.\,J. {Schlegel}},  \textsc{C.~{Bebek}},  \textsc{H.~{Heetderks}},
  \textsc{S.~{Ho}},  \textsc{M.~{Lampton}},  \textsc{M.~{Levi}},
  \textsc{N.~{Mostek}},  \textsc{N.~{Padmanabhan}},  \textsc{S.~{Perlmutter}}
  \etal{} \jr{ArXiv e-prints}(April) (2009).


\bibitem{Kazin:10}
 \textsc{E.\,A. {Kazin}},  \textsc{M.\,R. {Blanton}},
  \textsc{R.~{Scoccimarro}},  \textsc{C.\,K. {McBride}},  \textsc{A.\,A.
  {Berlind}},  \textsc{N.\,A. {Bahcall}},  \textsc{J.~{Brinkmann}},
  \textsc{P.~{Czarapata}},  \textsc{J.\,A. {Frieman}} \etal{} \jr{\apj}
  \textbf{710}(February), 1444--1461 (2010).


\bibitem{Percival:10}
 \textsc{W.\,J. {Percival}},  \textsc{B.\,A. {Reid}},  \textsc{D.\,J.
  {Eisenstein}},  \textsc{N.\,A. {Bahcall}},  \textsc{T.~{Budavari}},
  \textsc{J.\,A. {Frieman}},  \textsc{M.~{Fukugita}},  \textsc{J.\,E. {Gunn}}
  \etal{} \jr{\mnras} \textbf{401}(February), 2148--2168 (2010).


\bibitem{Dawson:13}
 \textsc{K.\,S. {Dawson}},  \textsc{D.\,J. {Schlegel}},  \textsc{C.\,P. {Ahn}},
   \textsc{S.\,F. {Anderson}},  \textsc{{\'E}.~{Aubourg}},
  \textsc{S.~{Bailey}},  \textsc{R.\,H. {Barkhouser}},  \textsc{J.\,E.
  {Bautista}},  \textsc{A.~{Beifiori}} \etal{} \jr{\aj} \textbf{145}(January),
  10 (2013).


\bibitem{Komatsu:11}
 \textsc{E.~{Komatsu}},  \textsc{K.\,M. {Smith}},  \textsc{J.~{Dunkley}}
  \etal{} \jr{\apjs} \textbf{192}(February), 18--+ (2011).


\bibitem{Padmanabhan:12a}
 \textsc{N.~{Padmanabhan}},  \textsc{X.~{Xu}},  \textsc{D.\,J. {Eisenstein}},
  \textsc{R.~{Scalzo}},  \textsc{A.\,J. {Cuesta}},  \textsc{K.\,T. {Mehta}},
  and  \textsc{E.~{Kazin}} \jr{\mnras} \textbf{427}(December), 2132--2145
  (2012).


\bibitem{Drinkwater:10}
 \textsc{M.\,J. {Drinkwater}},  \textsc{R.\,J. {Jurek}},  \textsc{C.~{Blake}},
  \textsc{D.~{Woods}},  \textsc{K.\,A. {Pimbblet}},  \textsc{K.~{Glazebrook}},
  \textsc{R.~{Sharp}},  \textsc{M.\,B. {Pracy}} \etal{} \jr{\mnras}
  \textbf{401}(January), 1429--1452 (2010).


\bibitem{Parkinson:12}
 \textsc{D.~{Parkinson}},  \textsc{S.~{Riemer-S{\o}rensen}},
  \textsc{C.~{Blake}},  \textsc{G.\,B. {Poole}},  \textsc{T.\,M. {Davis}},
  \textsc{S.~{Brough}},  \textsc{M.~{Colless}},  \textsc{C.~{Contreras}}
  \etal{} \jr{\prd} \textbf{86}(10), 103518 (2012).


\bibitem{Riess:09}
 \textsc{A.\,G. {Riess}},  \textsc{L.~{Macri}},  \textsc{S.~{Casertano}}
  \etal{} \jr{\apj} \textbf{699}(July), 539--563 (2009).


\bibitem{Jones:04}
 \textsc{D.\,H. {Jones}},  \textsc{W.~{Saunders}},  \textsc{M.~{Colless}},
  \textsc{M.\,A. {Read}},  \textsc{Q.\,A. {Parker}},  \textsc{F.\,G. {Watson}},
   \textsc{L.\,A. {Campbell}},  \textsc{D.~{Burkey}} \etal{} \jr{\mnras}
  \textbf{355}(December), 747--763 (2004).


\bibitem{Johnston:08}
 \textsc{S.~{Johnston}},  \textsc{R.~{Taylor}},  \textsc{M.~{Bailes}},
  \textsc{N.~{Bartel}},  \textsc{C.~{Baugh}} \etal{} \jr{Experimental
  Astronomy} \textbf{22}(December), 151--273 (2008).


\bibitem{Deboer:09}
 \textsc{D.\,R. {Deboer}},  \textsc{R.\,G. {Gough}},  \textsc{J.\,D. {Bunton}},
   \textsc{T.\,J. {Cornwell}},  \textsc{R.\,J. {Beresford}} \etal{} \jr{IEEE
  Proceedings} \textbf{97}(August), 1507--1521 (2009).


\bibitem{Duffy:12c}
 \textsc{A.\,R. {Duffy}},  \textsc{M.\,J. {Meyer}},
  \textsc{L.~{Staveley-Smith}},  \textsc{M.~{Bernyk}},  \textsc{D.\,J.
  {Croton}},  \textsc{B.\,S. {Koribalski}},  \textsc{D.~{Gerstmann}},  and
  \textsc{S.~{Westerlund}} \jr{\mnras} \textbf{426}(November), 3385--3402
  (2012).


\bibitem{Abdalla:12}
 \textsc{F.~{Abdalla}},  \textsc{J.~{Annis}},  \textsc{D.~{Bacon}},
  \textsc{S.~{Bridle}},  \textsc{F.~{Castander}},  \textsc{M.~{Colless}},
  \textsc{D.~{DePoy}},  \textsc{H.\,T. {Diehl}} \etal{} \jr{ArXiv
  e-prints}(September) (2012).


\bibitem{Levi:13}
 \textsc{M.~{Levi}},  \textsc{C.~{Bebek}},  \textsc{T.~{Beers}},
  \textsc{R.~{Blum}},  \textsc{R.~{Cahn}},  \textsc{D.~{Eisenstein}},
  \textsc{B.~{Flaugher}},  \textsc{K.~{Honscheid}} \etal{} \jr{ArXiv
  e-prints}(August) (2013).


\bibitem{FontRibera:13}
 \textsc{A.~{Font-Ribera}},  \textsc{P.~{McDonald}},  \textsc{N.~{Mostek}},
  \textsc{B.\,A. {Reid}},  \textsc{H.\,J. {Seo}},  and  \textsc{A.~{Slosar}}
  \jr{ArXiv e-prints}(August) (2013).


\bibitem{Meyer:04}
 \textsc{M.\,J. {Meyer}},  \textsc{M.\,A. {Zwaan}},  \textsc{R.\,L. {Webster}},
   \textsc{L.~{Staveley-Smith}},  \textsc{E.~{Ryan-Weber}},  \textsc{M.\,J.
  {Drinkwater}},  \textsc{D.\,G. {Barnes}},  \textsc{M.~{Howlett}} \etal{}
  \jr{\mnras} \textbf{350}(June), 1195--1209 (2004).


\bibitem{Zwaan:04}
 \textsc{M.\,A. {Zwaan}},  \textsc{M.\,J. {Meyer}},  \textsc{R.\,L. {Webster}},
   \textsc{L.~{Staveley-Smith}},  \textsc{M.\,J. {Drinkwater}},  \textsc{D.\,G.
  {Barnes}},  \textsc{R.~{Bhathal}},  \textsc{W.\,J.\,G. {de Blok}} \etal{}
  \jr{\mnras} \textbf{350}(June), 1210--1219 (2004).


\bibitem{Giovanelli:05}
 \textsc{R.~{Giovanelli}},  \textsc{M.\,P. {Haynes}},  \textsc{B.\,R. {Kent}},
  \textsc{P.~{Perillat}},  \textsc{A.~{Saintonge}},  \textsc{N.~{Brosch}},
  \textsc{B.~{Catinella}},  \textsc{G.\,L. {Hoffman}} \etal{} \jr{\aj}
  \textbf{130}(December), 2598--2612 (2005).


\bibitem{Nan:06}
 \textsc{R.~{Nan}} \jr{Science in China: Physics, Mechanics and Astronomy}
  \textbf{49}(March), 129--148 (2006).


\bibitem{Nan:11}
 \textsc{R.~{Nan}},  \textsc{D.~{Li}},  \textsc{C.~{Jin}},  \textsc{Q.~{Wang}},
   \textsc{L.~{Zhu}},  \textsc{W.~{Zhu}},  \textsc{H.~{Zhang}},
  \textsc{Y.~{Yue}},  and  \textsc{L.~{Qian}} \jr{International Journal of
  Modern Physics D} \textbf{20}, 989--1024 (2011).


\bibitem{Lonsdale:09}
 \textsc{C.\,J. {Lonsdale}},  \textsc{R.\,J. {Cappallo}},  \textsc{M.\,F.
  {Morales}},  \textsc{F.\,H. {Briggs}},  \textsc{L.~{Benkevitch}} \etal{}
  \jr{IEEE Proceedings} \textbf{97}(August), 1497--1506 (2009).


\bibitem{Booth:09}
 \textsc{C.\,M. {Booth}} and  \textsc{J.~{Schaye}} \jr{\mnras}
  \textbf{398}(September), 53--74 (2009).


\end{thebibliography}
\providecommand{\WileyBibTextsc}{}
\let\textsc\WileyBibTextsc
\providecommand{\othercit}{}
\providecommand{\jr}[1]{#1}
\providecommand{\etal}{~et~al.}

\end{document}